# High Performance Metallic Amorphous Magnetic Flake-based Magnetodielectric Inductors

Kun Qian[1], Alexander S. Sokolov[1], Qifan Li[1], Qifan Li[1], Chins Chinnasamy[2], Samuel Kernion[2], and Vincent G. Harris[1]**

[1]Center for Microwave Magnetic Materials and Integrated Circuits, Department of Electrical and Computer Engineering, Northeastern University, Boston, MA 02115, USA.
[2] Alloy development group, Carpenter Technology Corporation, Reading, PA 19601 USA.
** Fellow, IEEE

Abstract— Flake-shaped FeSi-based metallic amorphous alloy particles, having an aspect ratio as high as 175:1, were prepared by ball milling gas atomized amorphous powders of an effective diameter of 20 μm. The starting powder had a saturation magnetic flux density, $B_s$, of 1.5 T and a coercivity, $H_c$, of 94 A/m. The aspect ratio of the flakes, as well as their magnetic properties, were controlled by milling process parameters, such as duration, speed, and the type, mass and diameter of the milling balls. To minimize the oxidation of the charge, the powders were handled in an argon gas-purged glove box, milled in toluene, and subsequently dried in vacuo. Subsequently, soft magnetodielectric composites were prepared by suspending and aligning the FeSi-based powders in paraffin wax or epoxy resin. The composites were then pressed into toroids for measurements of their high frequency complex permeability by a vector network analyzer. The influence of the flakes' aspect ratio and volume loading fraction on the permeability of the composites were investigated. Results indicate that the composite permeability increases with the flakes' aspect ratio. For example, for a given loading factor of 30 vol.%, the composite permeability at 0.1 GHz nearly tripled and approached the value of 10 by increasing the aspect ratio of the FeSi-based inclusions from 1 (spheres) to greater than 175:1 (flakes).

## 1. INTRODUCTION

The rapid growth of semiconductor-based communication and computational platforms has made power and thermal management challenges a great concern, especially with the shifting of operational frequencies to higher bands. High performance inductors are required to provide substantial size reduction and effective thermal energy management. A path undertaken here includes the development of high saturation induction, high permeability, and soft magnetic flakes suspended in a low loss RF thermoset. Such flakes can be used in magnetodielectric composite inductors [1, 2] operating up to and beyond 200 MHz.

Applications with significant market impact include on-board inductors, power convertors, filters, etc. [3] and can be especially important to enable scaling of silicon-based voltage regulators, where miniaturization is driven by Moore's law. Fully Integrated Voltage Regulators (FIVR) will also experience greater efficiency and performance.

Iron powder cores are commercially available and perform with permeability values up to and beyond 100, but only at frequencies below 500 kHz [4]. High core losses stemming from eddy currents and related conduction losses limit higher frequency operation. A recent demonstration of ferrite cores with engineered grain boundary regions using insulating magnetic inclusions exhibit ultralow loss, high saturation magnetic flux density $B_s$, and sustained high permeability, and may operate beyond 1 MHz [5].

Based on calculations by Walser et al. [6], the use of shape anisotropy of soft magnetic particles can yield a maximum permeability at a given frequency, filling the frequency void where the maximum permeability could not be reached by the use of spherical powders. Furthermore, Walser and Kang [7] processed ferromagnetic flakes with sub-micrometer thicknesses and aspect ratios (length/thickness) between 10 and 100, showing promising microwave magnetodielectric properties. Recently, flake-shaped particles (Fe [8], FeCuNbSiB [9, 10], Sendust [11], FeSi [12], FeCo [13] and FeSiB [14]) with high aspect ratios and a thickness less than the skin depth have been shown to increase permeability [15], exceeding Snoek's limit [16], and reducing eddy current losses. Notwithstanding these results, there remains a void in available soft magnetic materials operating in VHF and UHF bands. The required soft magnetic material should possess a high $B_s$ to increase the operational frequency, low coercivity $H_c$ to decrease hysteretic losses, and high electrical resistivity to reduce conduction losses.

Corresponding author: Kun Qian (qian.kun1@husky.neu.edu).

The amorphous FeSiMe (where Me may be a combination of $Z_r$, B, P and Mn [17], henceforth FeSi-based alloy) soft magnetic powder used in this work was supplied by Carpenter Technology Corporation and was characterized as having ~20 μm average spherical particle size with a $B_s$ of 1.5 T, as compared to FeSiAl (1.2 T) [18] and NiZn ferrite (0.31T @ 10 kHz) [19]. The starting powder was produced by gas atomization. This process allows for consistent powder quality, and offers various advantages, such as high chemical purity and tolerance, and high sphericity.

Herein, amorphous FeSi-based flakes were processed by ball milling. It was demonstrated that the shape and aspect ratio of the flakes were controlled by adjusting the milling speed and duration, as well as the composition of the milling balls, that is, the type, mass, and sizes. The oblate spheroid with aspect ratio approaching 175:1 not only shows excellent magnetic properties, but also high permeability, and eddy current suppression above 2 GHz. The permeability of flake-based composites was found to be three times the value of spherical powder-based composites providing such composites as an attractive option for low loss applications in the 0.1 – 7 GHz frequency range.

## 2. EXPERIMENTAL PROCEDURE

### 2.1. Flake Preparation

The initial gas atomized FeSi-based amorphous alloy powder mostly consisted of spherical particles with a mean size ~20 μm (see Fig. 2(a)). Desired amounts of the starting powder were transferred to tungsten carbide jars, along with tungsten carbide milling balls, within a glove box purged with argon gas. Finally, toluene (≥ 99.5 % purity) was introduced to the jars as the milling medium, and the jars were hermetically sealed under argon gas pressure. The sealed jars were removed from the glove box and placed in a planetary ball mill (Across International PQ-N04) for milling.

To determine the optimal flake processing procedure, several critical parameters were systematically studied. The milling time was varied from 1 to 25 hours, while the rotation speed was kept constant at 500 rpm. The milling was executed using tungsten carbide balls with diameters ranging from 3 to 10 mm, as well as exploring several alternative combinations of different ball diameters. The ball-to-powder mass ratio varied from 2.6 to 10. A complete list of milling conditions is presented in Table 1. After milling, the jars were returned to the argon gas-purged glove box and unsealed. The powders were separated from the balls and washed three times in reagent-grade alcohol. Later, the powders were dried in vacuo for 1 hour within the exchange chamber of the glove box and eventually stored in the glove box for subsequent experiments. To reduce the coercivity, samples were annealed in a tube furnace for 1 hour in flowing argon gas at temperatures ranging from 300 to 450 °C. In contrast, one sample was annealed at 380 °C for 1 hour in a box furnace inside the glove box in an argon gas atmosphere. This final annealing was intended to reduce or eliminate strain introduced to the powders by the ballistic collisions experienced during milling. The reduction of strain is expected to reduce magnetoelastic anisotropy contributions to coercivity.

TABLE 1 HERE

### 2.2. Characterization and Measurements

The structure and morphology of samples were examined by room temperature θ-2θ powder x-ray diffraction (XRD, Rigaku Ultima IV, Cu Kα radiation (l=1.5406 Å)) and scanning electron microscopy (SEM, Hitachi S-4800), equipped with an energy-dispersive x-ray spectroscopy (EDS). The thickness and sizes of amorphous flakes were calculated using the Nano Measurer Plots software package. The room temperature magnetic properties of the starting materials and milled particles (powder density was presumed as 7.9 g/cm$^3$) were measured by a vibrating sample magnetometer (VSM, MicroSense Model EZ7) with a maximum applied magnetic field of 1.6 x 10$^6$ A/m. The soft magnetic composites were prepared by mixing the FeSi-based powder with paraffin wax or epoxy resin. The mixtures were pressed into toroids with an outer diameter of ~7 mm and an inner diameter of ~3 mm and a height of ~3 mm. The complex permeability of the toroids was measured using a coaxial airline and an Agilent E8363B vector network analyzer (VNA) from 0.1–18 GHz (see Fig. 1).

Fig.1 HERE

## 3. RESULTS and DISCUSSION

Fig. 2 illustrates the morphology of the particles before and after ball milling. Thickness, aspect ratio and static magnetic properties for samples prepared under different milling conditions are listed in Table 2. Comparing Figs. 2(b) through (f), the ball milling combination A2B3 (6 mm balls, 10:1 ball-to-powder weight ratio) deformed spheres principally into flakes after 8 hours of continuous milling. Therefore, this milling condition was found to be the most efficient. In contrast, ball milling condition A2B2 (6 mm balls, 5:1 ball-to-powder weight ratio) typically produced spherical particles after the same milling time. Smaller balls, e.g., 3 mm, showed little impact on the shape of the starting material. The aspect ratio of the flakes gradually increased with milling duration for most ball milling combinations. After 18 hours of ball milling, using the A2B3 combination, the resulting flakes achieved a mean thickness of 0.61 μm, and an aspect ratio exceeding 175:1 (Fig. 2(k)). Based on the comparison of sample morphologies in Figs. 2(h) and (i), ball milling produced diminishing returns after 18 hours.

FIG. 2 HERE

The flakes embedded in magnetodielectric composite inductor cores can be aligned by application of shear forces or an external magnetic field [20, 21] aligned along the in-plane direction of the flakes to disrupt eddy currents that manifest normal to the flake plane. If the flakes are much thinner than the skin depth at a given frequency, then the conduction losses can be dramatically reduced. Therefore, a thickness of about 0.61 μm effectively suppresses the eddy current loss of the composite inductor core because it is below the skin depth, which is about 1 μm at 1 GHz for iron as reported by Li et al. [22].

TABLE 2 HERE

Eddy current loss is the primary loss mechanism contributing to magnetodielectric composites at high frequencies. The other major source is the hysteretic loss that is reflected by the coercivity. The starting powder had a coercivity of 93.95 A/m. The coercivity gradually increased with milling duration and eventually reached 3105.10 A/m after 20 hours of ball milling, as the milling inevitably increased the residual stress within the flakes (Table 2). The broad diffraction peak appearing in the XRD pattern (Fig. 3(a)) indicated that the flakes remained amorphous even after milling durations of 20 hours. The saturation magnetic flux density of the material decreased by about 11 % after the powder was milled for 20 hours (Table 2).

FIG. 3 HERE

To reduce the stress and consequently the coercivity, the flakes were subsequently annealed for 1 hour at temperatures between 300-450 °C. Comparing the samples before (Fig. 2(h)) and after annealing (Figs. 2(j–l)), SEM results indicate that the influence of annealing on the size of flakes is negligible and the surface becomes smoother after annealing. As shown in Table 2, $H_c$ and $B_s$ initially improved with temperature and reached 923.57 A/m and 1.48 T at 380 °C, which is attributed to the release of stress during post-annealing. The lowest $H_c$ reported in the literature for comparable flakes is three times higher [23]. Magnetic properties gradually declined after annealing at higher temperatures. For example, $H_c$ increased dramatically after annealing at 450 °C because the amorphous flakes began to crystallize. XRD analysis (Fig. 3(a)) confirmed the partial crystallization of these samples. As identified in Fig. 3(a), the prominent diffraction peaks near 45º, 65º and 82 corresponds to the (110), (200) and (220) planes of α-Fe (Si) phase (PDF #184766).

Despite all precautions, we anticipated that flake processing would result in some level of oxidation of the starting material [24]. The oxidation is partially responsible for decreased $B_s$ measured after in milled samples [25]. EDX analyses (Fig. 3(b)) revealed that the oxygen content increased from 2.04 wt.% to 6.56 wt.% after milling the starting amorphous powders for 20 hours. At the same time, contamination caused by the tungsten carbide jar and balls used in milling could not be detected. The rate of oxidation was significantly reduced by degassing the oxygen in the vacuum chamber and annealing in the argon atmosphere of the glove box. This is consistent with the fact that the $B_s$ and lowest $H_c$ among the annealed samples occurred after a 380 °C annealing within the glove box.

As discussed above, the optimal experimental conditions included milling the powder with 6 mm diameter balls at ball-to-powder mass ratio of 10:1 in toluene for 18 hours and post annealing at 380 °C for 1 hour in the box furnace inside the argon gas-purged glove box. The $B_s$ and $H_c$ values of post-annealed flakes were 1.48 A/m and 598.73 A/m, respectively, as shown in the hysteresis loops of Fig. 4. The flakes of different sizes were segregated by sieving into categories of < 25 μm and > 45 μm. Figs. 2(k) and (l) illustrate SEM micrographs of sieved powders with flake dimensions corresponding to the < 25 μm and > 45 μm. It can be seen that most flakes were effectively segregated according to their diameter, and the thickness was mostly below 1 μm.

Fig.4 HERE

The complex permeability spectra were measured for composites prepared with the original spherical powders, as well as milled flakes. The composites had different loading factors and contained sieved flakes of different aspect ratios. As expected, the permeability of the > 45 μm flake composites, which reached 8.2 at 100 MHz, was almost three times higher than that of the composites consisting of spheres at a 30 vol.% loading (Fig.5(a)). The permeability of the > 45 μm flake composites was also higher than the permeability of the < 25 μm flake composites which reached 6.2. Additionally, the frequency corresponding to the peak of the magnetic loss tangent (tan $\delta = \mu'' / \mu'$) decreased with the increase of the flake aspect ratio in the 10-12 GHz range. The complex permeabilities of the > 45 μm flake composites prepared with epoxy resin and paraffin at 30 vol.% were similar but showed markedly different loss behavior. According to Walser et al.'s calculations [6], the maximum susceptibility $\chi_0'$ can be expressed as follows:

$$\omega_r^2 = [\gamma H_k + \gamma 4\pi M_S (D_\perp - D_e)][\gamma H_k + \gamma 4\pi M_S (D_h - D_e)] \quad (1),$$

$$\chi_0' = 4\pi M_S/[H_k + 4\pi M_S(D_h - D_e)] \quad (2),$$

$$\chi_0' f_r^2 = (2.8)^2 4\pi M_S[H_k + 4\pi M_S(D_\perp - D_e)] \quad (3),$$

where $\omega_r$ (represents the resonant frequency) $= 2\pi f_r$, $\gamma/2\pi = 2.8$ MHz/Oe, $H_k$ represents the anisotropy field, $D_e$ represents the demagnetization factor along the easy axis, $D_h$ represents the demagnetization factor along the microwave field direction and $D_\perp$ represents the demagnetization factor along the direction perpendicular to the easy axis and the microwave field direction. Equations (1)-(3) readily explain the increase of the permeability and decrease in the resonant frequency for flakes ($D_e \approx 1$, $D \approx D_h \approx 0$) and spheres ($D_e = D_\perp = D_h = 1/3$). The effective anisotropy field can be defined as the summation of anisotropy caused by the magnetoelastic coupling with the internal stress and shape anisotropy energy [26].

Fig.5 HERE

The eddy current losses are related to the diameter ($d$) of particles and the electric conductivity ($\sigma$) following (4) [27]:

$$\mu'' = \frac{2}{3}\pi\mu_0(\mu')^2 d^2 f\sigma \quad (4),$$

where $\mu_0$ is the vacuum permeability. If eddy current losses are the only source of magnetic losses, the value of $\mu''(\mu)^{-2}f^{-1}$ in (4) should be a constant with the change of frequency $f$. This is called the skin-effect criterion. Fig. 5(b) illustrates the value of $\mu''(\mu)^{-2}f^{-1}$ for the composite toroids containing 30 vol.% starting spherical power, flakes < 25 μm and flakes > 45 μm over the 0.1 – 18 GHz frequency range. As shown in Fig. 5(b), the values of the raw powder are close to 0.1 ns over 1 – 18 GHz. Thus, for the spherical composite, the magnetic loss mainly results from the eddy current over 1 – 18 GHz. Substituting the constant 0.1 ns and average diameter $d = 20$ μm into (4), the calculated electric conductivity is $9.6 \times 10^4$ S/m. For the flakes, the values are not constant over the frequency range, which can be apparent from the curve for flakes > 45 μm in Fig. 5(b). This means that the magnetic loss in the flakes is also related to the natural resonant loss. Because the change of the value of $\mu''(\mu)^{-2}f^{-1}$ for flakes > 45 μm in the frequency range is much greater than that for flakes < 25 μm, larger aspect ratio has a great influence of the natural resonance. This is also consistent with (3).

Comparing Figs. 5(c) and (d), the < 25 μm flake-based composites at a 40 vol.% loading achieved a permeability of 10, which is similar to that of the sphere-based composite at a 60 vol.% loading. Moreover, the flake-based composite maintained low loss tangents of below 0.5 at frequencies below 1 GHz (Fig. 5(f)), whereas magnetic loss tangents of the sphere-based composites increase rapidly with frequency (Fig. 5(e)). It is noteworthy that the permeability does not increase linearly with loading factor. The highest loading factor for flake composites achieved was 40 vol.%, which is limited by conduction losses.

## 4. CONCLUSION

FeSi-based alloy metallic amorphous spherical powders exhibit a high saturation magnetic flux density of 1.5 T, as well as low coercivity at 93.95 A/m. Such powders can be used in magnetodielectric composite inductor cores addressing the materials void in VHF and UHF frequency bands. In order to achieve the desired balance between high $B_s$ and low core loss properties, the spherical particles were converted to amorphous metallic flakes by ball milling. The flakes resulting from optimal processing possessed high aspect ratios in excess of 175: 1 and superior magnetic properties. The best results were achieved by continuous milling for 18 hours using 6 mm balls at 500 rpm. The final mean thickness of the flakes equals 0.61 μm and is sufficiently small to effectively inhibit eddy currents. The flakes were post-annealed at 380 °C for 1 hour in an argon gas environment to achieve a $B_s$ of 1.48 T and coercivity of 598.73 A/m. The flake composites showed higher permeability than original spherical powder composites at a 30 vol.% loading in paraffin (e.g. $\mu_{>45\,\mu m\,flake} = 8.2$, $\mu_{<25\,\mu m\,flake} = 6.2$ and $\mu_{original\,powder} = 3$ at 100 MHz). From the loading factor studies, the effective permeability of < 25 μm flake composites at 40 vol.% loading reached 10 at 100 MHz, similar to that of 60 vol.% loading of sphere composites. In the future work, we believe better insulation of flakes will further improve the performance of flake composites.

## ACKNOWLEDGMENT

This work was supported by the Rogers Corporation.

Table 1. Ball Milling Conditions for FeSi-based Alloy

| Level[b] | A: Stainless steel ball compositions | B: Ball-to-powder weight ratio |
|---|---|---|
| 1 | $\phi$[a]3 mm | 2.6:1 |
| 2 | $\phi$6 mm | 5:1 |
| 3 | $\phi$3 and $\phi$6 mm | 10:1 |
| 4 | $\phi$6 and $\phi$10 mm | - |

a $\phi$ represents the diameter.

b Level represents different milling condition combinations from ball diameter and ball-to-powder weight ratio. S (AxBy) means that the sample was milled using tungsten carbide ball composition as level x and ball-to-powder ratio as level y. For example, S (A1B1) represents the following milling condition: 3 mm diameter tungsten carbide balls and the ball-to-powder ratio of 2.6:1.

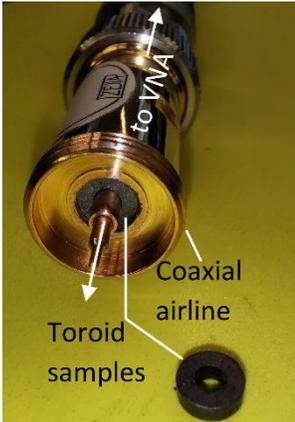

Fig.1. Image of complex permeability measurement setup.

Table 2. Flake Thickness and Length, Saturation Magnetic Flux Density and Coercivity of the As-Milled and Annealed FeSi-based Samples

| Sample | Milling time (h), annealing temp. (°C) | Flake thickness (μm) | Flake length (μm) | Saturation magnetic flux density (T) | Coercivity (A/m) |
|---|---|---|---|---|---|
| raw material | 0 | - | 1.9-58.6 | 1.50 | 93.95 |
| S (A4B1) | 4 | 1.9-20.3 | 17.2-35.4 | 1.49 | 2388.54 |
| | 8 | 0.6-26.4 | 4.1-52.3 | 1.49 | 2547.77 |
| | 8, 400 | 1.9-20.3 | 17.2-35.4 | 1.48 | 55.73 |
| S (A2B1) | 4 | 0.6-21.4 | 3.6-49.9 | 1.49 | 2547.77 |
| | 8 | 0.5-28.1 | 2.6-31.1 | -[b] | - |
| S (A2B2) | 8 | 1.2-26.8 | 8.7-46.3 | - | - |
| S (A1B1) | 4 | 0.8-27.1 | 5.3-53.3 | - | - |
| S (A3B3) | 10 | 0.1-22.7 | 3.4-36.2 | - | - |
| | 16 | 0.1-21.1 | 3.2-58.6 | - | - |
| | 25 | 0.1-20.5 | 2.1-33.9 | - | - |
| S (A2B3) | 8 | 0.8-15.2 | 4.5-38.6 | - | - |
| | 16 | 0.3-14.6 | 3.0-44.5 | - | - |
| | 18 | 0.1-1.3 | 2.7-88.6 | 1.47 | 1542.20 |
| | 18, 380 | | | 1.48 | 598.73 |
| | 20 | 0.1-1.6 | 2.3-62.1 | 1.34 | 3105.10 |
| | 20, 300[a] | | | 1.35 | 1496.82 |
| | 20, 380 | | | 1.48 | 923.57 |
| | 20, 400 | | | 1.34 | 1496.82 |
| | 20, 450 | | | 1.32 | 8208.60 |

a The annealed flakes were the flakes priory milled in toluene using A2B3 milling condition at 500 rpm for 20 hours.
b Samples without good flaky shape results did not measured or listed static magnetic properties.

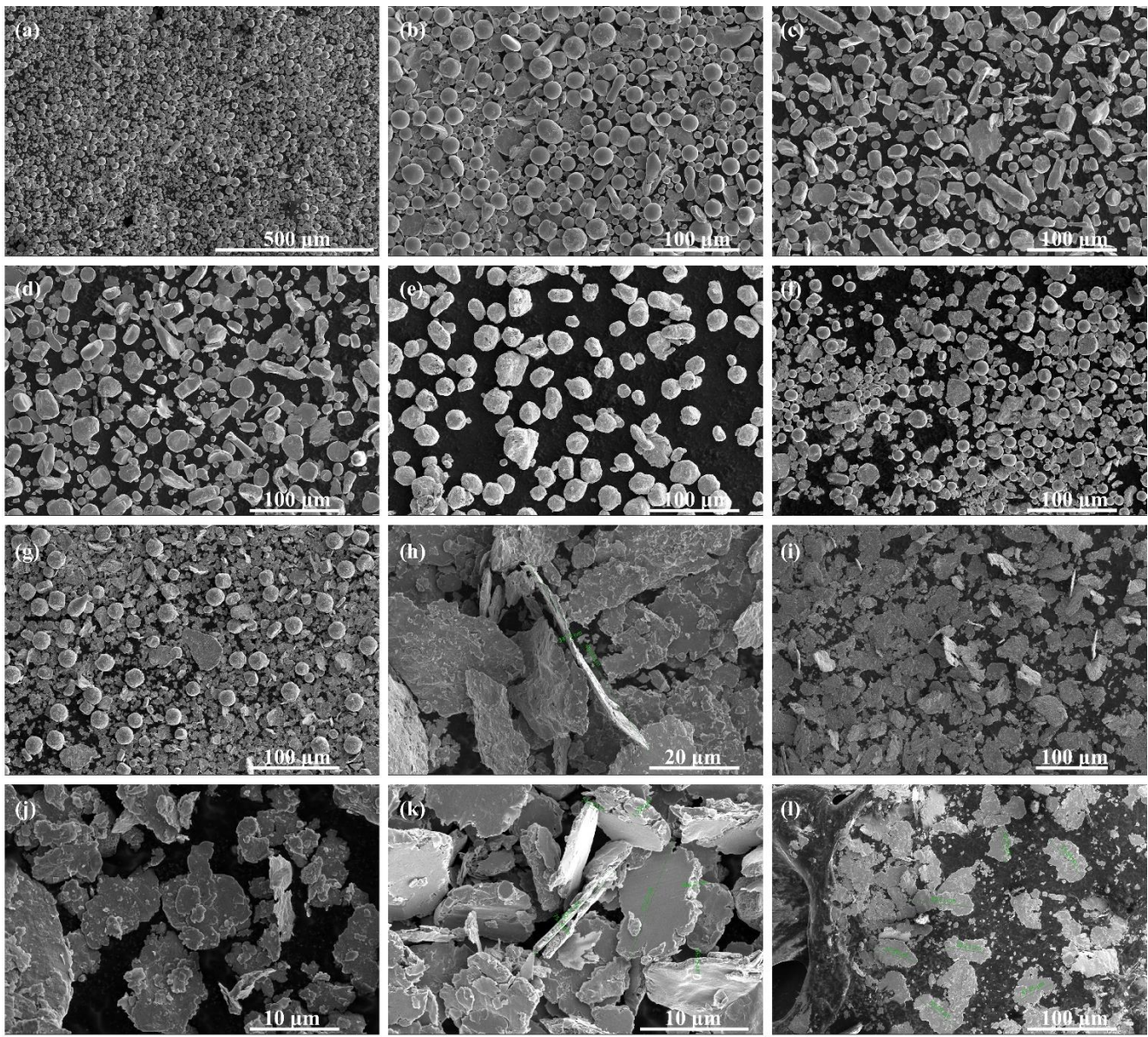

Fig. 2. Morphology of samples. (a) a SEM image of raw amorphous gas atomized FeSi-based powder. (b) (c), (d), (e) and (f) SEM images of powder milled using S (A1B1), S (A4B1), S (A2B1), S (A2B2) and S (A2B3) milling combinations for 8 hours, respectively. (g), (h) and (i) Morphology of powder milled at S (A3B3) condition for 19 hours, S (A2B3) for 18 hours and 20 hours, respectively. (j), (k) and (l) SEM images of particles at S (A2B3) condition milled for 18 hours and post-annealed at 380 °C for 1 hour, and then sieved to the length under 25 μm and above 45 μm.

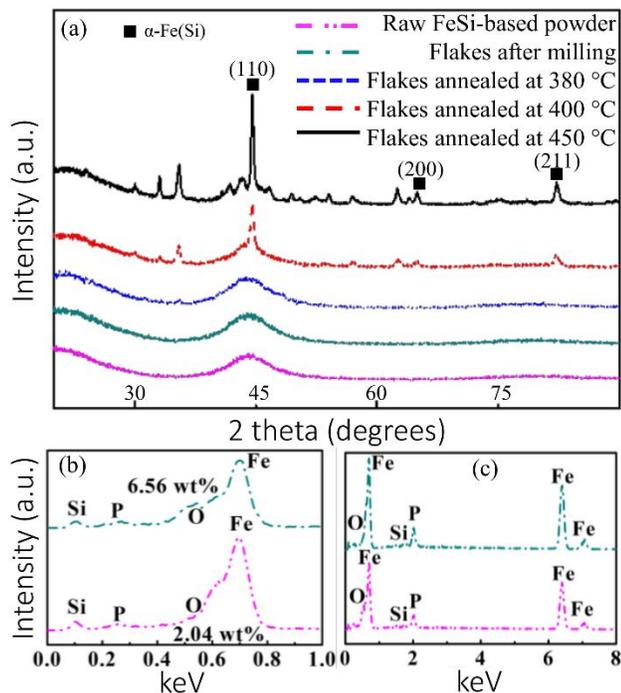

Fig. 3. (a) XRD patterns of original FeSi-based powder, the powder milled for 20 hours using 6 mm diameter tungsten carbide balls in Toluene with ball–to–powder weight ratio of 10 (A2B3) and the as-milled powder with post-annealing at 380, 400 and 450 °C for 1 h, respectively. (b) and (c) Energy dispersive X-ray (EDX) analysis of FeSi-based powder before and after ball milling for 20 hours using 6 mm diameter tungsten carbide balls with ball-to-powder weight ratio of 10.

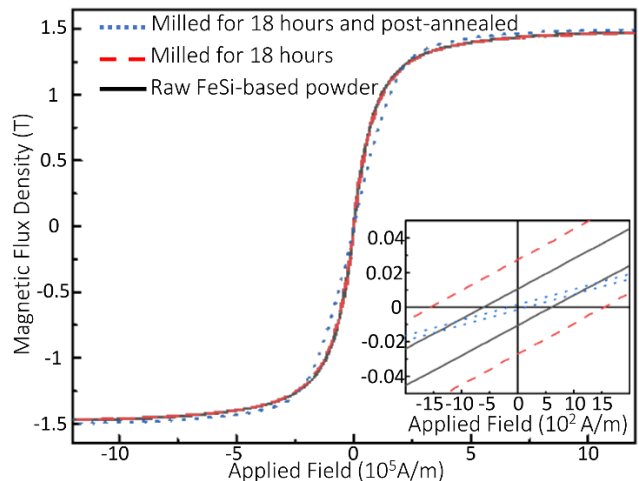

Fig.4. Hysteresis loops of original FeSi-based powder, flakes milled for 18 hours using 6 diameter balls at the ball-to-powder weight ratio as 10 and as-milled flakes with post-annealing in argon atmosphere at 380 °C for 1 hour, respectively. The insect shows the corresponding coercivity regions.

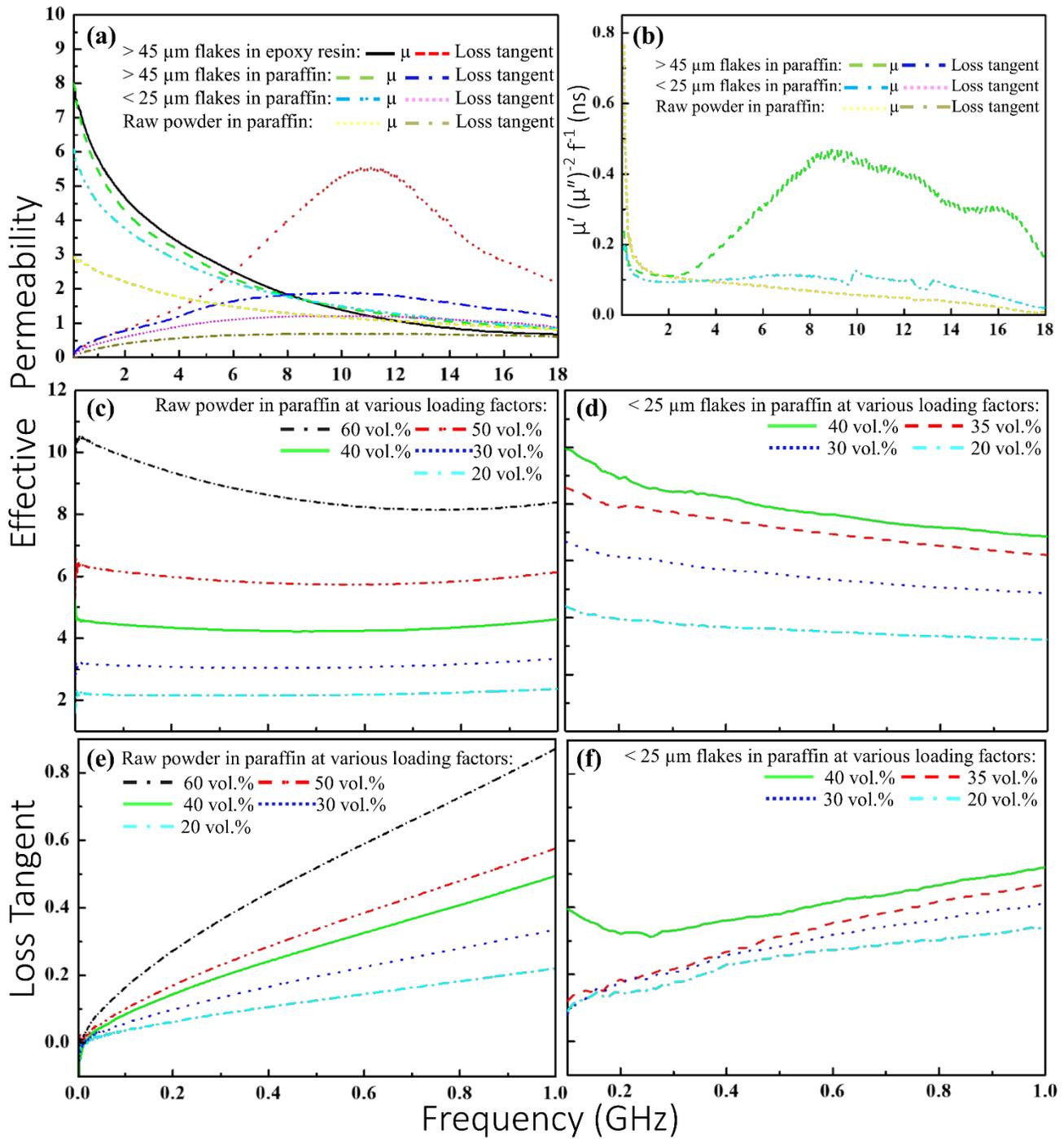

Fig. 5(a) Frequency dependence of effective permeability of original FeSi-based powder (20 μm mean diameter) and flakes in different sizes (under 25 μm and above 45 μm) at 30 vol.% in paraffin. (b) Value of $\mu''(\mu)^{-2}f^{-1}$ for the composite toroids made by the 30 vol.% raw spherical power, flakes under 25 μm and flakes above 45 μm vs frequency. (c) and (d) Frequency dependence of effective permeability of original powder and under 25 μm flakes in paraffin at different loading factors. (e) and (f) Frequency dependence of loss tangents of original powder and under 25 μm flakes in paraffin at different loading factors.